\documentclass[10pt]{article}
\usepackage{moriond,epsfig}

\def\be{\begin{equation}}
\def\ee{\end{equation}}
\def\bea{\begin{eqnarray}}
\def\eea{\end{eqnarray}}

\begin{document}
\vspace*{4cm}
\title{DARK MATTER PRIMORDIAL BLACK HOLES AND INFLATION MODELS}

\author{MANUEL DREES, ENCIEH ERFANI}

\address{Physikalisches Institut and Bethe Center for Theoretical Physics, 
 Universit\"{a}t Bonn\\ Nussallee 12, 53115 Bonn, Germany}

\maketitle\abstracts
{A broad range of single field models of inflation are analyzed in light of all relevant recent cosmological data, checking whether they can lead to the formation of long--lived Primordial Black Holes (PBHs) as candidate for dark matter. To that end we calculate the spectral index of the power spectrum of primordial perturbations as well as its first and second derivatives. PBH formation is possible only if the spectral index $n_S(k_0)$ increases significantly at small scales. Since current data indicate that the first derivative $\alpha_S$ of the spectral index is negative at the pivot scale, PBH formation is only possible in the presence of a sizable and positive second derivative (``running of the running'') $\beta_S$. Among the three small--field and five large--field models we analyze, only the ``running--mass'' model allows PBH formation, for a narrow range of parameters.}

\section{Introduction}

\paragraph{}
Inflation dynamically resolves many cosmological puzzles of the big bang model. On the other hand, the generation of a spectrum of primordial fluctuations in the early universe is a crucial ingredient of all inflationary models. These fluctuations can explain the generation of all (classical) inhomogeneities that can be seen in our universe. In addition, each inflationary scenario makes accurate predictions to discriminate between the various candidate models. One such prediction is the possible formation of PBHs. For this generation mechanism to be efficient, one typically needs a ``blue'' spectrum. In this way, one can hope that the density contrast is sufficiently large that the resulting PBH production is significant and can be used as a powerful constraint on the spectrum of inflationary primordial fluctuations. On the other hand, since the production of PBHs take place on scales much smaller than those probed by the CMB anisotropy and LSS formation, they are also a unique cosmological probe.

For single--field inflation models, the relevant parameter space for distinguishing among models is defined by the scalar spectral index $n_S$, the ratio of tensor to scalar fluctuations $r$, the running of the scalar spectral index $\alpha_S$ and we introduce a new parameter as ``running of running of the spectral index'', $\beta_S$. The goal of this paper is to make use of the recent observational bounds derived from the combined data of WMAP7 data, Baryon Acoustic Oscillations, $H_0$, South Pole Telescope and \textit{Clusters} \cite{WMAP7} to discriminate among the wide range of inflationary models, by checking if they can produce high density fluctuations at the scale relevant for long--lived PBHs.

\section{Primordial Black Holes and Inflation Models}
PBHs are black holes that result from the collapse of density fluctuations in the early universe where a lower threshold for the amplitude of such homogeneities is $\delta_{\rm th}\approx1/3$ at the time of radiation domination (RD). We have focused in our study \cite{second,encieh} on PBHs which form in the RD era after inflation and we also considered the standard case of PBHs formation, which applies to scales which have left the horizon at the end of inflation. We only consider gaussian and spherically symmetric perturbations and we use Press--Schechter formalism.
We study the possibility of PBH formation in two different categories of inflation models; small--field models and large--field models which their potentials are given as following:
\subsection{Small--field models}
\begin{eqnarray}
V(\phi)&=&V_0\left[ 1-\left( \frac{\phi}{\mu}\right) ^p \right]  \, ,\\
V(\phi)&=&V_0+\frac{1}{2}m_\phi^2(\phi)\phi^2\, , \quad {\rm Running-mass\quad inflation}\\
V(\phi)&=&V_0+\frac{\Lambda_3^{p+4}}{\phi^p}+...\, ,
\end{eqnarray}
\subsection{Large--field models}
\begin{eqnarray}
V(\phi)&=&\Lambda^4 \left( \frac {\phi}{\mu} \right) ^p\, ,\quad \rm{chaotic\quad inflation}\\
V(\phi)&=&\Lambda^4 e^{\left( \phi/\mu\right) ^p}\, ,\\
V(\phi)&=&V_0\left( 1-e^{-q\phi/M_{\rm P}} \right )\, ,\\
V(\phi)&=&\Lambda^4\left[1+\cos\left(\frac{\phi}{f} \right)  \right]\, ,\quad \rm{Natural\quad inflation}\\
V(\phi)&=&V_0\left[1+\frac{2}{\pi}\rm {arctan}\left(\frac{\phi}{\mu} \right) \right]\, , 
\end{eqnarray}
\section{Summary and Conclusions}
The formation of PBHs with mass larger than $10^{15}$ g, whose lifetime exceeds the age of the Universe, will be produced at sufficient abundance to form the cold Dark Matter if the spectral index at scale $k_{\rm PBH}$ is about $1.37$ for the threshold value $\delta_{\rm th}=1/3$. This spectral index is much above the value measured at much larger length scales in the CMB. PBH formation therefore requires significant positive running of the spectral index when $k$ is increased. We compared this with the values of the spectral index and its running derived from current data on large scale structure. At the pivot scale of the data set one finds $n_S(k_{\rm pivot}) = 0.9751$ as central value. The first derivative $\alpha_S(k_0)$ would then need to exceed $0.020$ if it alone were
responsible for the required increase of the spectral index; this is more than $3\,\sigma$ above the current central value of this quantity. However, the second derivative (the ``running of the running'') of the spectral index is currently only very weakly constrained. We showed in a model--independent analysis that this easily allows values of $n(k_{\rm PBH})$ large enough for PBH formation, even if the first derivative of the spectral index is negative at CMB scales. We applied this formalism \cite{second} to a wide class of inflationary models, under the constraints imposed by the data. We classified the inflation models in small--field and large--field models. We have shown that only one small--field model, the running--mass model \cite{encieh}, allows sizable positive running of running of the spectral index, and is thus a good candidate for long--lived PBHs
formation. In contrast, all the large--field models we studied predict small or negative values for the second derivative of the spectral index, and thus predict negligible PBH formation due to the collapse of overdense regions seeded during inflation.


\section*{References}
\begin{thebibliography}{99}
\bibitem{WMAP7} 
E.~Komatsu \textit{et. al.}, \textit{Astrophys.~J.~Suppl.} \textbf{192} (2011) 18; R.~Keisler \textit{et. al.}, \textit{Astrophys.~J.} \textbf{743} (2011) 28; W.~J.~Percival {\it et al.}, {\it MNRAS} {\bf 401} (2010) 2148; A.~G.~Riess {\it et al.}, {\it ApJ} {\bf 730} (2011) 119; A.~Vikhlinin \textit{et. al.}, \textit{ApJ} \textbf{692} (2009) 1060.

\bibitem{second}
M.~Drees and E.~Erfani, \textit{Primordial Black Holes in Single--Field Inflation Models}, \textit{JCAP} \textbf{1201} (2012) 035.

\bibitem{encieh}
M.~Drees and E.~Erfani, \textit{Running--Mass Inflation Models and Primordial Black Holes}, \textit{JCAP} \textbf{1104} (2011) 005.

\end{thebibliography}
\end{document}